\documentclass[superscriptaddress,prl,twocolumn]{revtex4}
\usepackage{graphicx}
\usepackage{amsmath}
\usepackage{amsfonts}
\usepackage{amssymb}

\begin{document}

\newcommand{\E}{\mathbb{E}}
\newcommand{\V}{\mathbb{V}}
\newcommand{\M}{\mathbb{M}}
\def\be{\begin{equation}}
\def\ee{\end{equation}}
\def\bea{\begin{eqnarray}}
\def\eea{\end{eqnarray}}
\newcommand{\avg}[1]{\langle{#1}\rangle}
\newcommand{\ve}[1]{\vec{#1}}

\title{Non-equilibrium mean-field theories on scale-free networks}

\author{Fabio Caccioli}
\affiliation{SISSA, via Beirut 4, I-34014 Trieste, Italy}
\affiliation{Istituto Nazionale di Fisica Nucleare, sezione di Trieste, Italy}
\author{Luca Dall'Asta}
\affiliation{Abdus Salam International Center for Theoretical Physics, Strada
  Costiera 11, 34014, Trieste, Italy}


\begin{abstract}
Many non-equilibrium processes on scale-free networks present anomalous critical behavior that is not explained by standard mean-field theories. We propose a systematic method to derive stochastic equations for mean-field order parameters that implicitly account for the degree heterogeneity.   The method is used to correctly predict the dynamical critical behavior of some binary spin models and reaction-diffusion processes. The validity of our non-equilibrium theory is furtherly supported by showing its relation with the generalized Landau theory of equilibrium critical phenomena on networks.

\end{abstract}

\maketitle

The discovery that many real systems can be represented, at some large-scale level, as complex networks with a disordered topology and an highly heterogeneous structure, has recently motivated a genuine interest for studying the properties of  dynamical processes occurring on top of them \cite{BBV08}.
The primary statistical feature that seems to have a strong influence on dynamical properties  is the existence of a very heterogeneous spectrum of local interaction patterns, reflected at the topological level into a very broad degree distribution.
A deeper understanding of its influence on the dynamics is obtained isolating this feature from the remaining complexity of real networks by considering ensembles of heterogeneous random graphs with power-law degree distribution  \cite{B80}.

Despite the local tree-like topology, random graphs present long loops ensuring that the average distance between any two points of the system scales at most logarithmically with the system's size \cite{CL02}. Therefore, physical properties on random graphs are not explicitly affected by spatial fluctuations.
Naively, this suggests that,  in the absence of additional  disorder and frustration, the general behavior of equilibrium and non-equilibrium  properties on the ensemble of uncorrelated random graphs should be of  mean-field (MF) type.
Nevertheless, analytical and numerical studies of statistical mechanics models defined on scale-free networks have revealed that fat-tails in the degree distribution can generate non-trivial critical behaviors in correspondence of continuous phase transitions \cite{DGM08}.
In networks with degree distribution $P(k) \sim k^{-\gamma}$, all moments $z_p \equiv \langle k^p \rangle = \sum_{k} k^p P(k)$ with $p > \gamma-1$ diverge with the system's size $N$; therefore scale-free networks with $2< \gamma <3$ present a finite average connectivity but uncontrolled degree fluctuations. On infinite networks, diverging degree fluctuations can prevent the occurrence of phase transitions,  and whenever they exist the
position of critical points and critical exponents explicitly depend on the value of $\gamma$.

The standard technique to investigate these phenomena on random graphs is the {\em heterogeneous mean-field theory} (HMF) \cite{DGM08},
namely by deriving a system of coupled mean-field equations for degree-dependent quantities. The approach is based on the {\em annealed network approximation}, in which network's topological characterization is strictly statistical, given in terms of the degree distribution $P(k)$ and, possibly, multipoint degree correlations.
Once the system of equations is solved, one computes the average global behavior in a self-consistent way to obtain the $\gamma$-dependent anomalous MF critical behavior.
A complementary approach proposed by Goltsev et al. \cite{GDM03} consists in assuming some reasonable form for the effective free-energy functional of the system and deriving from it the critical behavior of the corresponding MF order parameter. This method  generalizes the mean-field Landau-Ginzburg theory to take into account the topological heterogeneity of the system and provides a general conceptual framework to understand equilibrium critical phenomena on networks.
A similar general theory for non-equilibrium processes around critical points is still lacking, even if
an explicit dependence of the critical scaling behavior on the degree distribution's exponent $\gamma$  has been observed in several non-equilibrium models by means of the HMF theory and numerical simulations \cite{PV02,SR05,CPS06,CBPS05,BaCPS08,KJI06,BoCPS08}.
A main progress was recently achieved by showing that, like in equilibrium, the finite-size scaling does not depend only on the system's size $N$, but it may also depend on the upper cut-off $k_{c}$ of the degree distribution \cite{CPS08,HHP07}.
The non-equilibrium critical dynamics is therefore crucially influenced by the tail of the degree distribution and by the behavior of $k_{c}(N)$, particularly in the region $2 \leq \gamma \leq 3$.

In this Letter we provide a general mean-field approach taking into account the anomalous effects
introduced by degree heterogeneities, that could be applied to many classes of non-equilibrium processes in
a simple and systematic way, generalizing the equilibrium Landau-Ginzburg theory proposed by
Goltsev et al. \cite{GDM03}. We will first expose the general theory, for uncorrelated random graphs in the
annealed network approximation, then its validity is verified by direct application to three prototypical models
of non-equilibrium dynamics on networks: the voter model, the Ising-Glauber dynamics and the reaction-diffusion
processes. We show that degree heterogeneity affects the non-equilibrium  dynamics in different ways depending
whether or not the system's size $N$ and the  cut-off $k_{c}$ are finite. In networks with
a finite maximum degree, we eventually recover the mean-field scaling laws by tuning the control parameters
sufficiently close to criticality. On the contrary, in networks with a cut-off rapidly diverging with the system
size, the dynamics in the thermodynamic limit is governed by universal scaling laws that depend explicitly on the
exponent $\gamma$.

{\em General Theory -- \ }  In a general dynamical process on network, the
 microscopic dynamics is described by variables $\{x_i\}$ defined on the nodes $i = 1, \dots, N$; we consider their degree-dependent mean-field averages $X_k = \sum_{i: k_i=k} x_i / \sum_i \delta_{k_i,k}$. The corresponding HMF rate equations are of the form $\dot{X}_k = f(X_k, X)$, where $X$ is a global average quantity to be determined in a self-consistent way averaging over the degree distribution.  \\
In some cases the dynamics of $X$ is slow compared to the microscopic time-scales. This is evident $i$) when $X$ is a conserved quantity (e.g. in the voter model), but it is expected to be correct also $ii$) when the microscopic dynamics is driven by the behavior of high-degree nodes (e.g. for reaction-diffusion systems and Glauber dynamics) \footnote{Indeed the number of high-degree nodes is much smaller than $N$, therefore $X$ is averaged over a much larger set of values than $X_k$, and its increments are smaller on the average.}.
If $X$ is slow with respect to the relevant microscopic dynamics, we can use a quasi-static approximation imposing that $X_k$ rapidly relaxes to a quasi-stationary state that depends parametrically on the global quantity $X(t)$. The condition $\dot{X}_k \simeq 0$ allows to express $X_k=g(k,X(t))$ as a function of $k$ and $X(t)$. \\
The next step consists in deriving a stochastic equation for the slow variable $X(t)$. The mean
$\mathbb{E}[\Delta X]$ and the variance $\mathbb{V}[\Delta X]$ of the variation of $X$ during a unit time
interval can be easily computed. More precisely, one has to consider the variations of $X$ that are due to
single microscopic update events on nodes with degree $k$, then replace the expression $X_k=g(k,X(t))$ and
average over the degree distribution. At long times, close to the critical point, $X$ is small, so we can expand
these terms in powers of $X$ obtaining a Langevin equation of the type  $\dot{X}(t) = a X + b X^2 + \dots +
\sqrt{(a' X+ b' X^2 + \dots)/N } \eta(t)$, where $\eta(t)$ is a gaussian noise. Because of the expansion
of $g(k,X)$ for small $X$,  all coefficients are functions of the moments $z_p$ of the degree distribution.
In the $N \to \infty$ limit, noise terms will drop out but some of the other coefficients $a,b, \dots$ might diverge.\\
One can use an action formalism to verify that a Langevin equation with diverging contributions in the
thermodynamic limit is physically unreasonable. The probability of a trajectory $\{X(t),\hat{X}(t)\}$ is given
by  $P \propto \int \mathcal{D}[X,\hat{X}] e^{- N S[X,\hat{X}]}$, with action $S[X,\hat{X}] =\int dt
[\hat{X}\dot{X} - \hat{X} (a X + b X^2 + \dots) + \hat{X}^2 (a' X+ b' X^2 + \dots)]$.  As long as the network
has finite average degree ($\gamma \geq 2$), the action $S$ should be finite, whereas the
diverging coefficients contain terms of order $z_p \sim k_c^{p+1-\gamma} \sim
N^{\frac{p+1-\gamma}{\omega}}$ (for $k_c \sim N^{1/\omega}$). 
As we will directly verify later for several models, a physical action is recovered when assuming  the existence of some $X$-dependent
upper cut-off $k^{*}(X)$ such that above this value the dependence of $X_k$ on $k$ is weak and  does not generate
divergences. In other words, close to the critical points, the system should adjust itself in order to eliminate
unphysical divergences, with a mechanism similar to the one assumed by Goltsev et al. in their phenomenological
theory of equilibrium phase transitions on networks \cite{GDM03}.

In the following, the validity of this general method is demonstrated discussing three applications to relevant
non-equilibrium processes on scale-free networks.

{\em Voter Model -- \ } In the voter model (VM) \cite{L99}, each node of the system is endowed with a binary variable (opinion). At each time step, a randomly chosen node selects a neighbor at random and assumes its opinion. If we call $q_{k}(t)$ the probability that a node of degree $k$ has opinion $1$, the corresponding rate equation (HMF) is $\dot{q}_{k}(t) = Q(t) - q_{k}(t)$, where $Q(t) = \sum_{k} k P(k) q_k(t)/z_1$  is the probability to select a neighboring node with opinion $1$. According to the previously exposed theory, we identify  $X(t)=Q(t)$ as a good descriptor of the global dynamics of the system.  
Mean and variance of the variations $\Delta Q$ can be easily computed on uncorrelated random graphs in the MF approximation \cite{VE08}. $Q$ is a conserved quantity, so $\mathbb{E}[\Delta Q] = 0$ and we arrive to  the fluctuations-driven Langevin equation
\begin{equation}
\dot{Q}(t) = \sqrt{\frac{z_2}{N z_1^2} Q(t)\left[1-Q(t)\right]}\eta(t)
\end{equation}
where $\eta(t)$ is a delta correlated white noise. Processes with a global conserved quantity have a trivial
infinite-size limit, but it is still interesting to study the effects of the degree heterogeneity on the
finite-size scaling of $\tau$, the characteristic time needed to reach an absorbing state. Here $Q(t)$ performs
a kind of random-walk and reaches one of the absorbing states ($Q = 0,1$) in a characteristic time $\tau\propto
N z_1^2/z_2$.  For a power-law network with cutoff $k_{c} \sim N^{1/\omega}$, we have  $\tau\sim N$ for
$\gamma>3$, $\tau\sim N/\log N$ for $\gamma=3$,  and $\tau\sim N^{(\omega-3+\gamma)/\omega}$ for $2<\gamma<3$,
in agreement with the results by Sood and Redner \cite{SR05}.
Studying directly the stochastic  evolution of $q(t) = \sum_k P(k) q_{k}(t)$ would have led incorrectly to the mean-field result $\tau \propto N$ also for $\gamma <3$.\\
Our approach can be applied to the reverse VM (or invasion process) as well, for which the degree-dependent rate equations suggest to study the global quantity  $R(t) = \sum_k P(k) z_1 q_k(t) / k$. As previously $Q(t)$, now $R(t)$ is a conserved quantity, and the corresponding Langevin equation gives $\tau \propto  N$ \cite{C05}.

{\em Glauber-Ising dynamics -- \ }  With respect to the Voter model, the Glauber-Ising dynamics has non-zero drift and
allows to directly verify the relation between our method and the equilibrium theory of critical
phenomena on networks. Consider the finite-temperature dynamics on a general graph \cite{G63}: on each node $i$, there is
a binary spin  $\sigma_{i} = \pm1$ and spin-flipping events occur with rate $[ 1+ \sigma_i \tanh(\beta
\sum_{j\in v(i)} \sigma_j)]/2$ ($v(i)$ are the neighbors of $i$).  In the annealed network approximation, one can easily derive the rate equation
for the degree-dependent magnetization $\dot{m}_k = -m_k + \tanh(\beta k h(t))$, where $h = \sum_k P(k) h_k =
\sum_k k P(k) m_{k}/z_1$ is the average local field.
Note that in the quasi-static approximation $m_k \simeq \tanh(\beta k h)$, that close to the critical point can be approximated by  $m_k \propto k h$ for $k \ll k^* = 1/\beta h$ and $m_k \approx 1$ for $k \gg k^* = 1/\beta h$.  Hence, $h_k \propto k^2 h/z_1$ for $k \ll k^*$ and $h_k \approx 1$ for $k \gg k^*$. \\
A detailed study of the stochastic  evolution equation for $h(t)$ will be presented elsewhere \cite{inprep}. Here we consider the mean-field rate equation surviving in the infinite size limit, i.e.
\begin{equation}\label{lf}
\frac{d}{dt} h(t) = -h(t) + \sum_k \frac{k P(k)}{z_1} \tanh{\left[ \beta k h(t) \right]} .
\end{equation}
Expanding the hyperbolic tangent for small $h \ll 1$, we have $\dot{h} \simeq - h + \beta h \frac{\langle k^2 \rangle}{z_1} - \beta^3 h^3 \frac{\langle k^4 \rangle}{3 z_1}$. The behavior of the system depends on the relation between the real cut-off $k_c$ and the dynamic cut-off $k^*(t) \propto 1/h(t)$.  At the critical point $h(t)$ relaxes to zero, therefore there is a typical time $t_c$ such that $k^*(t_c)=k_c$, after that any finite system behaves in a mean-field way, except for some anomalous $N$-dependent prefactors (and the effect of the noise that we have neglected \cite{inprep}). This is true also for infinite systems with finite maximum degree ($k_c < \infty$).
In these systems the phase transition occurs at $\beta = \beta_c = z_1/z_2 \propto k_c^{\gamma-3}$ \cite{DGM08}.
The critical relaxation dynamics $m(t) \sim t^{-1/z}$ is given by solving $\dot{h} \propto - \frac{z_4}{z_2^3} h^3$ and assuming that 
close to the critical point $m = \sum_k P(k) m_k \simeq h$.
The standard MF exponent $z=2$ is recovered, but there is a cut-off dependent prefactor  that accelerates the dynamics.\\
On infinite networks with diverging degree cut-off, the sums over $k$ have to be performed with the modified
upper bound $k^{*} \propto 1/h$ to avoid unphysical divergences. Indeed for much larger degrees the Taylor
expansion itself is not correct, because the hyperbolic tangent is close to $1$. In other words, $g(k,h) =
\tanh(\beta k h)$ for the Glauber dynamics. Performing the averages with cut-off $k^*$ and neglecting higher
orders in $h$, we find $\dot{h}(t) \simeq -h + A(\beta h)^{\gamma -2}$ for  $2 < \gamma <3$ ($A$ is a constant),
whereas the second term gets a logarithmic correction $\sim A \beta h \log(const/h)$ for $\gamma = 3$. There is
no paramagnetic phase at finite temperature ($\beta_c = 0$) \cite{DGM08} and the critical dynamics is
exponentially fast.
 For $3 < \gamma <5$, the second moment is finite but $\langle k^4 \rangle \to \infty$, therefore the effective theory reads  $\dot{h}(t) = -h + A\beta h - B (\beta h)^{\gamma -2}$. The critical temperature is now finite and the critical dynamics presents the anomalous MF exponent $z = \gamma  -3$. The standard MF exponent $z=2$ is recovered only for $\gamma > 5$.\\
Our results corroborate the idea previously suggested in Ref.\cite{DGM08} that the non-equilibrium Ising model
on networks could be studied using the dissipative dynamics  $d m / dt = - \frac{\delta
\mathcal{F}_{\gamma}[m]}{\delta m}$ of soft spins under a generalized $\gamma$-dependent Landau-Ginzburg
potential $\mathcal{F}_{\gamma}$.

{\em Reaction-Diffusion processes -- \ }  In reaction-diffusion processes \cite{VK81}, particles are free to diffuse on a network and interact with a given rate when they are on the same site. We are interested in "bosonic" process, i.e. there are no restrictions on node's occupancy \cite{BaCPS08}.  \\
The natural degree-dependent quantity to study is the average number of particles $\rho_k$ in nodes of degree $k$. Each variable satisfies the rate equation $\dot{\rho}_{k}(t) = - \rho_k + k \rho/ z_1 + \Lambda[\rho_k]$, where the first two terms are due to diffusion in uncorrelated random graphs and the third is a reaction kernel that depends only on $\rho_k$.
For simplicity, we focus on the dynamics of branching-annihilating particles, i.e.  $2 A \xrightarrow{\lambda/2} 0$ and $A \xrightarrow{\mu} 2 A$, with reaction kernel $\Lambda[\rho_k]  = \mu  \rho_k - \lambda\rho_{k}^2$.\\
The global variable contained in the degree-dependent rate equations is the particle density $\rho$. Its relation with $\rho_k$ is obtained self-consistently by solving the rate equations in the quasi-static approximation,  $\rho_k(t) \approx \frac{\mu-1}{2\lambda}[1-\sqrt{1+4\lambda k \rho(t)/((\mu-1)^2 z_1)}]$. Now that we have $\rho_k = g(k,\rho(t))$, we develop $g$ for small values of $\rho$, close to the critical point $\mu_c = 0$. We find two regimes:
$\rho_{k}(t) \propto k \rho(t) / z_1$ for $k \ll k^* \propto z_1 / \rho(t)$ and $\rho_{k}(t) \propto \sqrt{k \rho(t)/z_1}$  for $k \gg k^* $. \\
Starting from the degree-dependent variations $\Delta \rho_k$ and averaging over $P(k)$, we find the Langevin equation
\begin{equation}
\dot{\rho}(t) = \mu  \rho -  \lambda \langle \rho_{k}^2 \rangle+ N^{-1/2}\sqrt{\mu \rho + 2 \lambda \langle
\rho_{k}^2 \rangle}\eta(t) .
\end{equation}
According to the above relation between $\rho_k$ and $\rho$, the average over $P(k)$ can be splitted in two integrals: one for $k< k^{*}$ and the other for $k>k^{*}$. In a finite network with maximum degree $k_c$,
\begin{equation}\label{kappastar}
\int P(k) \rho_{k}^2 dk \propto \left\{ \begin{array}{cc}   \rho^{\gamma-1} + const \times k_c^{2-\gamma} \rho , &    k_c > k^*  \\   k_c^{3-\gamma} \rho^2 , &  k_c < k^* . \end{array} \right.
\end{equation}
Close to the critical point, the particle density  is small, thus we expect that on finite networks and infinite networks with finite cut-off the system will eventually enter the mean-field regime ($k^{*} > k_c$).
The corresponding Langevin equation is
\begin{equation}\label{langevin}
\dot{\rho}(t) \simeq a  \rho - b' \rho^2 + \sqrt{(a \rho +  b' \rho^2 )/ N}\eta(t)
\end{equation}
in which the various constants have been absorbed in two effective reaction rates $a\propto \mu$ and $b' =2 b
k_c^{3-\gamma}$ ($b \propto \lambda$). Assuming $k_c \sim N^{1/\omega}$, the stationary density in the active
phase follows a mean-field scaling $\rho_{act} \propto A(N) \mu$, with a non-trivial prefactor $A(N) \sim
N^{-\frac{3-\gamma}{\omega}}$. For a generalized reaction $qA \to \emptyset$, we have $\rho_{act} \propto
\mu^{\frac{1}{q-1}} N^{-\frac{q+1-\gamma}{(q-1)\omega}} $. For finite $N$ the active state is metastable, that
is the system will eventually decay to the absorbing  state because of noise-induced fluctuations. The amplitude
of the leading noise term scales like $N^{-\frac{\gamma+\omega-3}{\omega}}$. In case of networks with structural
cut-off ($\omega = 2$), the noise is always small at large but finite $N$. If we prepare a system in the
metastable state ($\rho_{act} \sim 1/\sqrt{N}$) just above the critical point ($\mu > 0$),  it will decay to the
absorbing state in a time $ e^{\mathcal{O}(N^{(\gamma-1)/2})}$. But in a network with natural cut-off  $\omega =
\gamma-1$, the dynamical fluctuations become pathological in the limit $\gamma \to 2$:  the density of the
metastable state is just $\mathcal{O}(1/N)$ and it will decay to $\rho =0$ in a finite timescale. The problem of
large fluctuations induced by multiplicative noise in scale-free networks requires furtherly investigations
 \cite{inprep}.

In infinite networks with diverging degree cut-off $k_c$, at every time we have $k^{*}(t) < k_c$,  so using Eq.\ref{kappastar} we get the effective rate equation $\dot{\rho}(t) \simeq a \rho - b \rho^{\gamma -1}$.
The exponent $\gamma -1$ is universal, independent of the number $q$ of particles involved in the annihilation reaction.
In summary, the general effective mean-field equation for the critical behavior of reaction-diffusion processes on uncorrelated random networks with $P(k) \sim k^{-\gamma}$ ($\gamma > 2$) is
\begin{equation}\label{infinite}
\dot{\rho}(t) \simeq a \rho - b \rho^{\gamma -1} - c \rho^q ,
\end{equation}
where the second term dominates over the third for $\gamma-1 < q$ and viceversa. Logarithmic corrections appear for $\gamma-1 =q$.
This expression for two-particles annihilation, $q=2$, resembles the phenomenological theory proposed by Hong et al. \cite{HHP07}. However, our expression is strictly correct only in the infinite-size limit, when the stochastic term due to the granularity of the system is absent.  The addition of a multiplicative noise term, such as $\sqrt{\rho} \eta$ proposed in \cite{HHP07}, wherever justified, should not change the mean-field behavior obtained from Eq. \ref{infinite}.

A similar analysis could be performed for "fermionic" (or hard-core) particles as well (i.e. nodes are occupied by at most one particle) \cite{CBPS05,BaCPS08}.  For instance, the fermionic branching-annihilation model is described by a rate equation of the form
$\dot{\rho}_{k}(t) = -\rho_k(t)+k \rho /z_1 + \Lambda[\rho_k,\rho]$, with  $\Lambda[\rho_k,\rho] = \nu k[1-\rho_k(t)]\rho(t) /z_1 - 2 k \rho(t) \rho_k(t) /z_1$ ($\nu$ is an effective reaction rate) \cite{BaCPS08}. At fixed global density, we observe the scaling $\rho_{k}(t) \propto k \rho(t) /z_1$ for $k \ll k^{*} \propto z_1/\rho(t)$ and $\rho_{k}(t) \sim const. < 1$ otherwise. Hence previous calculations can be easily extended to fermionic models, obtaining qualitatively similar results \cite{inprep}. In particular, our analysis corroborates recent results by Bogu\~n\'a at al. \cite{BoCPS08} for the contact-process on networks.

{\em Conclusions --} We have proposed a general method to derive  improved mean-field theories for
non-equilibrium critical phenomena on heterogeneous networks. It allows to choose the correct mean-field order
parameter for many different models, and obtain the anomalous mean-field scaling laws in a systematic way for
both finite-size and infinite-size systems. This approach is simpler than the widely used HMF method and
represents  the non-equilibrium generalization of the equilibrium theory of critical phenomena on scale-free
networks \cite{GDM03}.  The method was successfully applied to the voter model, the Glauber dynamics and
reaction-diffusion processes. For all of them, the nature of critical behavior depends on the cut-off of the
degree distribution. For finite cut-offs, the system eventually recovers mean-field behavior close to the
critical point, whereas fluctuations may strongly affect single realizations of the process for very
heterogeneous networks \cite{inprep}. On the contrary, infinite networks with diverging degree cut-off present
an anomalous mean-field critical behavior, directly dependent on the exponent $\gamma$ of the degree
distribution.

\begin{acknowledgments}
We warmly thank A. Baronchelli, A. Barrat, C. Castellano, S. Franz and M. Marsili for useful discussions. F. C.
acknowledges the grant 2007JHLPEZ (MIUR).
\end{acknowledgments}

\end{document}